\documentclass[twocolumn,prl,showpacs,superscriptaddress]{revtex4}
\usepackage{graphicx}
\usepackage{amsmath}
\begin{document}

\title{Observation of minority spin character of the new electron
doped manganite La$_{0.7}$Ce$_{0.3}$MnO$_3$ from tunneling
magnetoresistance}

\author{C. Mitra}
\affiliation{Max Planck Institute for Chemical Physics of Solids,
N\"othnitzer Str. 40, 01187 Dresden, Germany}
\affiliation{Institut f\"ur Festk\"orper und Werkstofforschung
Dresden, Helmholtzstrasse 20, 01069 Dresden, Germany}
\author{P. Raychaudhuri}
\affiliation{School of Physics and Astronomy, University of
Birmingham, Edgbaston, Birmingham, B15 2TT, UK}
\author{K. D\"orr}
\affiliation{Institut f\"ur Festk\"orper und Werkstofforschung
Dresden, Helmholtzstrasse 20, 01069 Dresden, Germany}
\author{K.-H. M\"uller}
\affiliation{Institut f\"ur Festk\"orper und Werkstofforschung
Dresden, Helmholtzstrasse 20, 01069 Dresden, Germany}
\author{L. Schultz}
\affiliation{Institut f\"ur Festk\"orper und Werkstofforschung
Dresden, Helmholtzstrasse 20, 01069 Dresden, Germany}
\author{P.~M. Oppeneer}
\affiliation{Institut f\"ur Festk\"orper und Werkstofforschung
Dresden, Helmholtzstrasse 20, 01069 Dresden, Germany}
\author{S. Wirth}
\affiliation{Max Planck Institute for Chemical Physics of Solids,
N\"othnitzer Str. 40, 01187 Dresden, Germany}

\date{\today}

\begin{abstract}We report the magnetotransport characteristics of a
trilayer ferromagnetic tunnel junction build of an electron doped
manganite (La$_{0.7}$Ce$_{0.3}$MnO$_3$) and a hole doped manganite
(La$_{0.7}$Ca$_{0.3}$MnO$_3$). At low temperatures the junction
exhibits a large positive tunneling magnetoresistance (TMR),
irrespective of the bias voltage. At intermediate temperatures
below $T_C$ the sign of the TMR is dependent on the bias voltage
across the junction. The magnetoresistive characteristics of the
junction strongly suggest that La$_{0.7}$Ce$_{0.3}$MnO$_3$ is a
minority spin carrier ferromagnet with a high degree of spin
polarization, i.e. a transport half metal.
\end{abstract}
 \maketitle

There has been a lot of interest recently in the hole doped
rare-earth manganites, where the rare-earth in the insulating
parent compound is partially replaced by a divalent cation (such
as Ca, Ba, Sr, Pb etc.) \cite{coe}. Around 30\% hole doping most
of these compounds such as La$_{0.7}$Ca$_{0.3}$MnO$_3$,
La$_{0.7}$Ba$_{0.3}$MnO$_3$ have a half-metallic ferromagnetic
ground state with a large magnetoresistance (MR) associated with
the ferromagnetic transition temperature. Conversely, doping
electrons by substituting the rare-earth atom by tetravalent Ce
also drives the system into a ferromagnetic metallic ground state
\cite{man,ray,mit1}. Unlike the hole doped compound where
manganese is in a mixture of Mn$^{3+}$ and Mn$^{4+}$ valence
states the tetravalent Ce drives the compound in a mixture of
Mn$^{3+}$ and Mn$^{2+}$ valencies. The presence of
Mn$^{2+}$/Mn$^{3+}$ valencies induced by electron doping in
La$_{0.7}$Ce$_{0.3}$MnO$_3$ was confirmed recently \cite{xas}.
There exists an inherent symmetry between Mn$^{4+}$ and Mn$^{2+}$
as both are non Jahn-Teller ions, whereas Mn$^{3+}$ is a
Jahn-Teller ion. Moreover, La$_{0.7}$Ca$_{0.3}$MnO$_3$ and
La$_{0.7}$Ce$_{0.3}$MnO$_3$ both have a Curie temperature of $T_C
\approx$ 250 K.

Beyond the phenomenon of colossal MR, novel properties arising
from the interplay of spin, charge and orbital coupling and the
competition of closely related energy scales make the manganites
potential candidates for novel electronic devices. Devices
exhibiting both large positive and negative MR have been
fabricated by integrating doped manganites with other oxide
ferromagnets. More recently rectifying characteristics could be
demonstrated in
La$_{0.7}$Ce$_{0.3}$MnO$_3$/SrTiO$_3$(STO)/La$_{0.7}$Ca$_{0.3}$MnO$_3$
tunnel junctions \cite{mit2} and Nb-STO/(La,Ba)MnO$_3$ junctions
\cite{tan}. However, a detailed understanding of the spin
dependent electronic structure of these materials is important to
exploit the complete potential of these compounds.

\begin{figure}[tb]
\centering \includegraphics[width=8.0cm]{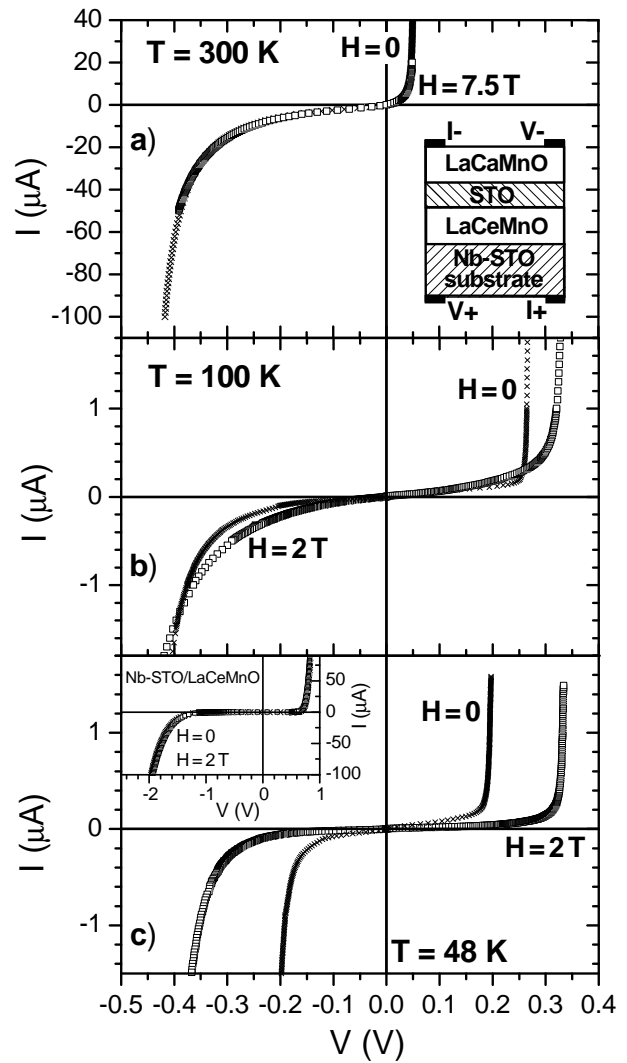}
\vspace*{-0.2cm} \caption{(a) The tunneling $I-V$ characteristics
of the La$_{0.7}$Ce$_{0.3}$MnO$_3$/STO/La$_{0.7}$Ca$_{0.3}$MnO$_3$
trilayer junction, taken at 300 K, in zero field and in an
in-plane field of 7.5 T. Inset: layout of the multilayer device.
(b) The same as before, taken at 100 K, in zero field and in a
field of 2 T. (c) The same as above, at 48 K. Inset: $I-V$ curve
of La$_{0.7}$Ce$_{0.3}$MnO$_3$ on Nb-STO at 48 K for comparison.}
\label{fig1}
\end{figure}

In this paper we report the magnetotransport properties of a
{La$_{0.7}$Ce$_{0.3}$MnO$_3$/STO($\approx\!
50\:\!$\AA)/La$_{0.7}$Ca$_{0.3}$MnO$_3$} tunnel junction grown
onto conducting 0.5\% Nb doped SrTiO$_3$ (Nb-STO) single
crystalline substrate through pulsed laser deposition, whose
fabrication has been reported earlier \cite{mit2}. The device
structure is illustrated in the inset of Fig.~\ref{fig1}(a). The
Nb-STO substrate acts as a conducting underlayer allowing us to
measure the transport properties of the tunnel junction with
current perpendicular to plane (CPP) geometry. In contrast to the
hole doped compounds not much work \cite{min1,min2} could so far
be done to understand the electronic structure of
La$_{0.7}$Ce$_{0.3}$MnO$_3$. A major experimental obstacle stems
from the fact that La$_{0.7}$Ce$_{0.3}$MnO$_3$ has so far been
synthesized in single phase only in thin film form through
high-energy pulsed laser deposition \cite{mit1}. This is a
restriction on many of the conventional probes and calls for
alternative ways to extract information regarding the spin
polarized electronic structure in these materials. Towards this
end we have used magnetotransport properties of tunnel junctions
comprising of two ferromagnetic electrodes separated by a thin
insulating tunnel barrier. Most importantly, we wanted to
establish whether this electron doped ferromagnet has minority
spin carriers (MISC) (where the magnetization is antiparallel to
the spin) or majority spin carriers (MASC) (magnetization parallel
to the spin) at the Fermi energy $E_F$. The resistance across such
a ferromagnetic tunnel junction (FTJ) depends on the relative
orientation of the spins at $E_F$ in the two electrodes. In zero
field the magnetizations of the two ferromagnetic electrodes in an
ideal FTJ are predominantly antiparallel owing to magnetostatic
interaction between the ferromagnetic layers with in-plane
magnetization. With the application of a magnetic field the
magnetizations become parallel. Thus, a tunnel junction where both
electrodes are either MISC or MASC ferromagnets exhibits negative
tunneling magnetoresistance (TMR). This is observed in manganite
tunnel junctions of identical materials separated by a thin STO
insulating barrier \cite{gup,sun}. In contrast, for a tunnel
junction with one MISC and one MASC ferromagnetic electrode the MR
is positive. This has experimentally been observed in
La$_{0.7}$Sr$_{0.3}$MnO$_3$/STO/Fe$_3$O$_4$ tunnel junctions
\cite{gho} showing large positive MR, where Fe$_3$O$_4$ is a MISC
half-metallic ferrimagnet and La$_{0.7}$Sr$_{0.3}$MnO$_3$ is the
MASC half-metallic ferromagnet. Here, we investigate the spin
character of La$_{0.7}$Ce$_{0.3}$MnO$_3$ at $E_F$ from the
magnetotransport properties of the tunnel junction using the MASC
ferromagnet La$_{0.7}$Ca$_{0.3}$MnO$_3$ as a spin analyzer.

Fig.~\ref{fig1}(a) shows the magnetic field dependence of the
current versus voltage ($I-V$) curve across the tunnel junction
measured in the CPP geometry at 300 K. We do not see a significant
TMR, as expected, since at this temperature both
La$_{0.7}$Ce$_{0.3}$MnO$_3$ and La$_{0.7}$Ca$_{0.3}$MnO$_3$ are
paramagnetic semiconductors and the device behaves like a
rectifying diode. However, in a field of 7.5 T the spin disorder
scattering of a single layer is reduced, and the in-plane MR is
quite large \cite{coe}. Fig.~\ref{fig1}(b) shows the magnetic
field dependence of the tunneling $I-V$ curve taken at 100 K, in
zero field and in a field of 2 T. The magnetic field dependence of
the $I-V$ curves (both positive and negative bias) clearly shows a
bias dependent MR, i.e., below a threshold bias voltage (or
current), we find a negative MR and above this a positive MR
prevails. The field dependence of the $I-V$ curves taken at 48 K
[Fig.~\ref{fig1}(c)] does not show any bias dependence, exhibiting
a positive MR  at all voltages. In order to rule out any
significant influence of the Nb-STO/La$_{0.7}$Ce$_{0.3}$MnO$_3$
bottom junction on the results of the tunnel junction we also
deposited single layers of La$_{0.7}$Ce$_{0.3}$MnO$_3$ (and
La$_{0.7}$Ca$_{0.3}$MnO$_3$) of identical thickness upon Nb-STO.
As a main result, the MR of the Nb-STO/La$_{0.7}$Ce$_{0.3}$MnO$_3$
bottom junction is several orders of magnitude smaller compared to
that of the FTJ and is always negative. Moreover, the $I-V$ curve
of the former is highly asymmetric [inset Fig.~\ref{fig1}(c)] down
to the lowest temperature.

\begin{figure}[tb]
\centering
\includegraphics[width=8.1cm]{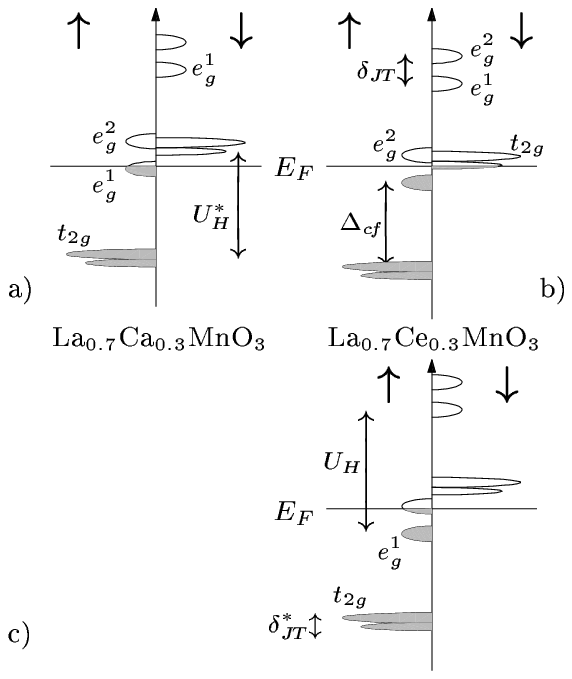}
\caption{Schematic diagram of spin dependent (large arrows)
density of states of (a) the Ca and (b) and (c) the Ce-doped
compounds at low temperature. For the Ce-doped compound, panel (b)
[panel (c)] depicts the case of the level $t_{2g}\!\!\downarrow$
being energetically lower [higher] than $e_g^2\!\!\uparrow$
resulting in a MISC [MASC] state. Our tunneling experiments
indicate MISC at $E_F$ for La$_{0.7}$Ce$_{0.3}$MnO$_3$. For
tunneling, panels (a) and (b) in the diagram correspond to the
high field case, i.e., aligned magnetizations within the two
ferromagnetic layers and hence, increased resistance due to
opposite spin states at $E_F$. (See text for definitions of the
energies.)} \label{fig2}
\end{figure}

First we concentrate on the $I-V$ characteristics of the tunnel
junction at the lowest temperature (48 K). Here, the positive TMR
is intriguing. In hole doped manganites, three of the Mn-3d
electrons form the localized $t_{2g}$ band. The remaining
electrons occupy the conducting $e_g$ band which is energetically
higher than the $t_{2g}$ in a (nearly) cubic crystal field. The
crystal field splitting energy is estimated to be $\Delta_{cf}$
$\approx$ 1.8 eV \cite{coe}. The $e_g$ band splits further due to
Jahn-Teller distortion into two sub-bands, $e_{g}^{1}$ and
$e_{g}^{2}$, which are separated by the Jahn-Teller splitting
energy, $\delta_{JT}$ $\approx$ 1.2 eV \cite{coe,sat}. Jahn-Teller
distortion also causes a splitting of the $t_{2g}$ band by
$\delta_{JT}^*$. Moreover, Hund's rule coupling removes the spin
degeneracy in the ferromagnetic state. The resulting separation of
the spin-up ($e_g\!\!\uparrow$) and spin-down
($e_g\!\!\downarrow$) bands is denoted by $U_H$ whereas the
separation of the $t_{2g}\!\!\uparrow$ and $t_{2g}\!\!\downarrow$
bands is labeled $U_H^*$. The Mn-3d spin dependent density of
states (DOS) of La$_{0.7}$Ca$_{0.3}$MnO$_3$ at low temperature is
sketched in Fig.~\ref{fig2}(a). The bandwidth of the conducting
$e_g$ sub-bands is of the order of 1 eV \cite{coe}. $U_H$ is
estimated from band structure calculations to be of the order of 2
eV in the undoped compound---a value that may decrease with doping
in the hole doped compound \cite{oki}. In hole doped manganites
like La$_{0.7}$Ca$_{0.3}$MnO$_3$, which have 0.7 electrons in the
$e_g$ band, the conduction electrons predominantly occupy the
lowest sub-band $e_{g}^{1}\!\!\uparrow$ and the hole doped
manganite is consequently a MASC ferromagnet [Fig.~\ref{fig2}(a)].

In La$_{0.7}$Ce$_{0.3}$MnO$_3$, there is clear evidence for
electron doping on the Mn-site in {\em single} layer epitaxial
films, e.g., from x-ray absorption spectroscopy \cite{xas}. Hence,
the $e_{g}^{1}\!\!\uparrow$ sub-band is completely filled. For the
remaining additional (doped) electrons two scenarios are possible:
i) weak Hund's rule coupling $U_H^* < \Delta_{cf} + \delta_{JT}$
and ii) strong Hund's rule coupling $U_H^* > \Delta_{cf} +
\delta_{JT}$. In the first case, $t_{2g}\!\!\!\downarrow$ is
energetically lower than $e_{g}^{2}\!\!\uparrow$ and the former
will get partially filled [Fig.~\ref{fig2}(b)]. The compound will
be a MISC ferromagnet in an intermediate spin state. In the second
case, the remaining electrons will occupy the
$e_{g}^{2}\!\!\uparrow$ sub-band resulting in a MASC ferromagnet
in high spin state [Fig. \ref{fig2}(c)]. The observation of a
positive TMR in
La$_{0.7}$Ce$_{0.3}$MnO$_3$/STO/La$_{0.7}$Ca$_{0.3}$MnO$_3$ at low
temperature definitely favors the first scenario with antiparallel
spins at $E_F$ for fields high enough to align the magnetizations
within the two ferromagnetic layers [cf. Fig. \ref{fig2}(a) and
(b)]. This result is supported by the known energy values given
above if $U_H^* \approx U_H$ is assumed. Band structure
calculations \cite{sat} also predict the near-degeneracy of the
energy positions of the $e_g^2\!\!\uparrow$ and the
$t_{2g}\!\!\downarrow$ sub-bands.

The minority spin character and the related intermediate spin
state observed in La$_{0.7}$Ce$_{0.3}$MnO$_3$ are intriguing for
several reasons. First, due to the usually large on-site Hund's
rule coupling this is rarely observed in manganese compounds where
Mn is in the divalent state including compounds such as MnO.
Secondly, the transport and magnetic properties of the hole doped
manganites are conventionally understood from this large on-site
Hund's rule coupling in manganese and the electron-lattice
coupling arising from the Jahn-Teller effect in Mn$^{3+}$. The
ferromagnetic ground state in hole doped compounds like
La$_{0.7}$Ca$_{0.3}$MnO$_3$ is driven by the double exchange
interaction between Mn$^{3+}$ and Mn$^{4+}$. This, however, is
unlikely to be the dominant mechanism in
La$_{0.7}$Ce$_{0.3}$MnO$_3$ if the manganese is in intermediate
spin state. The striking similarity between the magnetic and
transport properties of La$_{0.7}$Ce$_{0.3}$MnO$_3$ and its hole
doped counterpart \cite{ray} raises questions anew regarding the
origin of colossal MR in these compounds. It should be pointed out
in this context that even in the hole doped compound the existence
of Mn$^{3+}$ and Mn$^{4+}$ in the pure high spin state has been
questioned recently. From point contact Andreev reflection
measurements \cite{nad} the existence of minority spin carriers in
the hole doped compound La$_{0.7}$Sr$_{0.3}$MnO$_3$ was concluded.
Studies on tunnel junctions using a superconducting electrode and
La$_{0.7}$Sr$_{0.3}$MnO$_3$ \cite{geb} also showed the
polarization $P$ $= (n_{\uparrow}(E_F) -
n_{\downarrow}(E_F))/(n_{\uparrow}(E_F) + n_{\downarrow}(E_F))$
($n_{\uparrow}$ and $n_{\downarrow}$ being the spin up and spin
down DOS) to be of the order of 72\%, which is much smaller than
the polarisation expected for a half-metal \cite{par}. Existing
band structure calculations \cite{pic} predict an even lower spin
polarization. These results indicate that the manganese is not in
the pure high spin state in both electron and hole doped
manganites with perovskite crystal structure. Therefore the
general explanation of transport and magnetic properties of
colossal MR manganites in terms of double exchange as the dominant
mechanism needs a closer examination regardless of their high
half-metallic transport character.

\begin{figure}[tb]
\centering
\includegraphics[width=7.4cm]{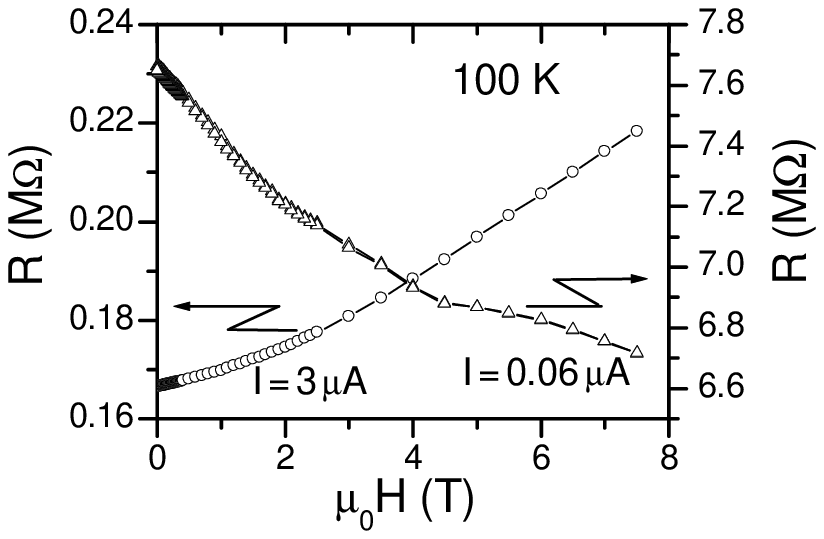} \vspace*{-0.4cm}
\caption{Tunneling resistance ($R$) vs. magnetic field ($H$) at
100 K for two different bias currents $I = 0.06 \,\mu$A and $3
\,\mu$A.} \label{fig3}
\end{figure}

Fig.~\ref{fig3} shows the resistance versus field ($R-H$) of the
tunnel junction at 100 K. At a very small bias current ($I = 0.06
\,\mu$A) the MR is negative while at a large current ($I = 3
\,\mu$A) the MR is positive. A description of the bias dependence
of the MR at 100 K [Fig.~\ref{fig1}(b)] and 200 K (not shown) is
beyond the ground state picture described in Fig.~\ref{fig2}.
Here, a detailed knowledge of the band structure and the
temperature evolution of the bands in La$_{0.7}$Ce$_{0.3}$MnO$_3$
would be required. Since at high bias the large electric field
across the insulating tunnel barrier shifts their Fermi energies
considerably with respect to each other, we first concentrate on
the temperature dependent TMR observed at low bias. The crossover
of the TMR from positive to negative at low bias suggests a change
in the dominant spin character of the ferromagnet at $E_F$ with
temperature. This could happen for two reasons. First, in
transition metal oxides the interplay between competing energy
scales can induce a crossover from one spin state to another with
temperature. For example, in LaCoO$_3$ where the crystal field
energy slightly exceeds the Hund's rule coupling energy, Co is in
a low spin, effectively non-magnetic state at low temperature.
Around 100 K the cobalt ions undergo a crossover from a low spin
state to an intermediate/high spin state accompanied by a change
in magnetic and electric properties \cite{Ima}. Since the
$t_{2g}\!\!\downarrow$ and $e_g^2\!\!\uparrow$ bands are at
similar energy levels (and may even overlap) in the manganites, a
temperature dependent crossover from an intermediate spin state to
a high spin state cannot be ruled out in
La$_{0.7}$Ce$_{0.3}$MnO$_3$. It may be caused, e.g., by decreasing
$\Delta_{cf}$ with increasing temperature. This needs to be
substantiated through other measurements. Secondly, even in the
absence of such a crossover in the Mn spin state the dominant spin
character of the carriers {\em at} $E_F$ in a ferromagnet can
change. For example, in a Stoner ferromagnet the exchange
splitting between spin up and spin down bands decreases with
increasing temperature, which changes the filling of the up and
down spin bands. Therefore, subtle features in the DOS as a
function of energy $E$ can modify the relative ratio of up ($e_g$)
and down ($t_{2g}$) spins at $E_F$ without changing the overall
spin state of the underlying Mn-ions.

We now focus on the bias dependence of the MR at 100 K and 200 K.
A bias voltage $V_b$ on a metallic tunnel junction shifts the
Fermi levels of the two electrodes by $eV_b$. The tunneling of
electrons across an insulating barrier, however, occurs at equal
energy levels. Hence, the tunneling probabilities for the
magnetization of the two electrodes parallel (antiparallel) to
each other, $T_{\uparrow\uparrow}$ ($T_{\uparrow\downarrow}$), are
given by
\begin{eqnarray*}
T_{\uparrow\uparrow} & \propto & n_{\uparrow}(E^+)\,
n'_{\uparrow}(E^-)\; +\; n_{\downarrow}(E^+)\,
n'_{\downarrow}(E^-) \; , \\ T_{\uparrow\downarrow} & \propto &
n_{\uparrow}(E^+)\, n'_{\downarrow}(E^-) \; +\;
n_{\downarrow}(E^+)\, n'_{\uparrow}(E^-)
\end{eqnarray*}
respectively, where $E^+ = E_F + \frac{eV_{b}}{2}$ and $E^- = E_F
- \frac{eV_{b}}{2}$. Here, $n_{\uparrow}(E)$ and
$n_{\downarrow}(E)$ are the spin up and spin down DOS in
La$_{0.7}$Ca$_{0.3}$MnO$_3$ and $n'_{\uparrow}(E)$ and
$n'_{\downarrow}(E)$ are the corresponding ones in
La$_{0.7}$Ce$_{0.3}$MnO$_3$. The relative magnitude of
$T_{\uparrow\uparrow}$ and $T_{\uparrow\downarrow}$ depend on the
details of the spin up and spin down DOS. At 100 K and 200 K, the
negative MR at low bias voltage suggests that
$T_{\uparrow\uparrow} > T_{\uparrow\downarrow}$ giving a negative
MR. However, at the same temperatures, at large $V_b$ the MR may
change sign if this inequality is altered owing to features in the
energy dependent DOS. Moreover, small changes in the relative
energies of the $t_{2g}\!\!\downarrow$ and $e_g^2\!\!\uparrow$
bands due to $V_b$ may change their occupancy which may result in
a bias dependent crossover. Here, a detailed understanding can
emerge only when the DOS in La$_{0.7}$Ce$_{0.3}$MnO$_3$ is known
in sufficient detail.

In summary, we have reported the magnetotransport properties of a
tunnel junction made of electron and hole doped manganites. The
observed minority spin carrier transport in
La$_{0.7}$Ce$_{0.3}$MnO$_3$ is of fundamental interest to
understand the interplay of the Hund's rule coupling energy with
other energy scales such as Jahn-Teller energy and crystal field
energy in doped manganites. We believe that this result will
motivate further studies of the exact electronic structure of both
electron and hole doped manganites and may open up an alternative
approach towards {\em spintronics}.

The authors would like to acknowledge Andy Mackenzie, F. Steglich,
S.K. Dhar and R. Pinto for helpful discussions and encouragement.
P.R. and C.M. thank the Leverhulme Trust and DFG (Grant No. SFB
422), respectively, for financial support.


\begin{thebibliography}{19}
\expandafter\ifx\csname
natexlab\endcsname\relax\def\natexlab#1{#1}\fi
\expandafter\ifx\csname bibnamefont\endcsname\relax
  \def\bibnamefont#1{#1}\fi
\expandafter\ifx\csname bibfnamefont\endcsname\relax
  \def\bibfnamefont#1{#1}\fi
\expandafter\ifx\csname citenamefont\endcsname\relax
  \def\citenamefont#1{#1}\fi
\expandafter\ifx\csname url\endcsname\relax
  \def\url#1{\texttt{#1}}\fi
\expandafter\ifx\csname
urlprefix\endcsname\relax\def\urlprefix{URL }\fi
\providecommand{\bibinfo}[2]{#2}
\providecommand{\eprint}[2][]{\url{#2}}

\bibitem[{\citenamefont{Coey et~al.}(1999)\citenamefont{Coey, Viret, and von
  Moln\'ar}}]{coe}
\bibinfo{author}{\bibfnamefont{J.}~\bibnamefont{Coey}},
  \bibinfo{author}{\bibfnamefont{M.}~\bibnamefont{Viret}}, \bibnamefont{and}
  \bibinfo{author}{\bibfnamefont{S.}~\bibnamefont{von Moln\'ar}},
  \bibinfo{journal}{Adv.\ Phys.} \textbf{\bibinfo{volume}{48}},
  \bibinfo{pages}{167} (\bibinfo{year}{1999}).

\bibitem[{\citenamefont{Mandal and Das}(1997)}]{man}
\bibinfo{author}{\bibfnamefont{P.}~\bibnamefont{Mandal}} \bibnamefont{and}
  \bibinfo{author}{\bibfnamefont{S.}~\bibnamefont{Das}},
  \bibinfo{journal}{Phys.\ Rev.\ B} \textbf{\bibinfo{volume}{56}},
  \bibinfo{pages}{15073} (\bibinfo{year}{1997}).

\bibitem[{\citenamefont{Raychaudhuri et~al.}(1999)}]{ray}
\bibinfo{author}{\bibfnamefont{P.}~\bibnamefont{Raychaudhuri}}
  \bibnamefont{et~al.}, \bibinfo{journal}{J.\ Appl.\ Phys.}
  \textbf{\bibinfo{volume}{86}}, \bibinfo{pages}{5718} (\bibinfo{year}{1999}).

\bibitem[{\citenamefont{Mitra et~al.}(2001{\natexlab{a}})\citenamefont{Mitra,
  Raychaudhuri, John, Dhar, Nigam, and Pinto}}]{mit1}
\bibinfo{author}{\bibfnamefont{C.}~\bibnamefont{Mitra}},
  \bibinfo{author}{\bibfnamefont{P.}~\bibnamefont{Raychaudhuri}},
  \bibinfo{author}{\bibfnamefont{J.}~\bibnamefont{John}},
  \bibinfo{author}{\bibfnamefont{S.~K.} \bibnamefont{Dhar}},
  \bibinfo{author}{\bibfnamefont{A.~K.} \bibnamefont{Nigam}}, \bibnamefont{and}
  \bibinfo{author}{\bibfnamefont{R.}~\bibnamefont{Pinto}},
  \bibinfo{journal}{J.\ Appl.\ Phys.} \textbf{\bibinfo{volume}{89}},
  \bibinfo{pages}{524} (\bibinfo{year}{2001}{\natexlab{a}}).

\bibitem[{\citenamefont{Mitra et~al.}({\natexlab{a}})}]{xas}
\bibinfo{author}{\bibfnamefont{C.}~\bibnamefont{Mitra}} \bibnamefont{et~al.},
  \bibinfo{note}{cond-mat/0206137}.

\bibitem[{\citenamefont{Mitra et~al.}(2001{\natexlab{b}})}]{mit2}
\bibinfo{author}{\bibfnamefont{C.}~\bibnamefont{Mitra}} \bibnamefont{et~al.},
  \bibinfo{journal}{Appl.\ Phys.\ Lett.} \textbf{\bibinfo{volume}{79}},
  \bibinfo{pages}{2408} (\bibinfo{year}{2001}{\natexlab{b}}).

\bibitem[{\citenamefont{Tanaka et~al.}(2002)\citenamefont{Tanaka, Zhang, and
  Kawai}}]{tan}
\bibinfo{author}{\bibfnamefont{H.}~\bibnamefont{Tanaka}},
  \bibinfo{author}{\bibfnamefont{J.}~\bibnamefont{Zhang}}, \bibnamefont{and}
  \bibinfo{author}{\bibfnamefont{T.}~\bibnamefont{Kawai}},
  \bibinfo{journal}{Phys.\ Rev.\ Lett.} \textbf{\bibinfo{volume}{88}},
  \bibinfo{pages}{027204} (\bibinfo{year}{2002}).

\bibitem[{\citenamefont{Min et~al.}(2001)\citenamefont{Min, Kwon, Lee, and
  Kang}}]{min1}
\bibinfo{author}{\bibfnamefont{B.~I.} \bibnamefont{Min}},
  \bibinfo{author}{\bibfnamefont{S.~K.} \bibnamefont{Kwon}},
  \bibinfo{author}{\bibfnamefont{B.~W.} \bibnamefont{Lee}}, \bibnamefont{and}
  \bibinfo{author}{\bibfnamefont{J.-S.} \bibnamefont{Kang}},
  \bibinfo{journal}{J.\ Electron\ Spectrosc.\ Relat.\ Phenom.}
  \textbf{\bibinfo{volume}{114}}, \bibinfo{pages}{801} (\bibinfo{year}{2001}).

\bibitem[{\citenamefont{Kang et~al.}(2001)\citenamefont{Kang, Kim, Lee, Olson,
  and Min}}]{min2}
\bibinfo{author}{\bibfnamefont{J.-S.} \bibnamefont{Kang}},
  \bibinfo{author}{\bibfnamefont{Y.~J.} \bibnamefont{Kim}},
  \bibinfo{author}{\bibfnamefont{B.~W.} \bibnamefont{Lee}},
  \bibinfo{author}{\bibfnamefont{C.~G.} \bibnamefont{Olson}}, \bibnamefont{and}
  \bibinfo{author}{\bibfnamefont{B.~I.} \bibnamefont{Min}},
  \bibinfo{journal}{J.\ Phys.:\ Condens.\ Matter}
  \textbf{\bibinfo{volume}{13}}, \bibinfo{pages}{3779} (\bibinfo{year}{2001}).

\bibitem[{\citenamefont{Gupta et~al.}(1996)}]{gup}
\bibinfo{author}{\bibfnamefont{A.}~\bibnamefont{Gupta}} \bibnamefont{et~al.},
  \bibinfo{journal}{Phys.\ Rev.\ B} \textbf{\bibinfo{volume}{54}},
  \bibinfo{pages}{R15629} (\bibinfo{year}{1996}).

\bibitem[{\citenamefont{Sun et~al.}(1996)}]{sun}
\bibinfo{author}{\bibfnamefont{J.~Z.} \bibnamefont{Sun}} \bibnamefont{et~al.},
  \bibinfo{journal}{Appl.\ Phys.\ Lett.} \textbf{\bibinfo{volume}{69}},
  \bibinfo{pages}{3266} (\bibinfo{year}{1996}).

\bibitem[{\citenamefont{Ghosh et~al.}(1998)}]{gho}
\bibinfo{author}{\bibfnamefont{K.}~\bibnamefont{Ghosh}} \bibnamefont{et~al.},
  \bibinfo{journal}{Appl.\ Phys.\ Lett.} \textbf{\bibinfo{volume}{73}},
  \bibinfo{pages}{689} (\bibinfo{year}{1998}).

\bibitem[{\citenamefont{Satpathy et~al.}(1996)\citenamefont{Satpathy, Popovic,
  and Vukajlovic}}]{sat}
\bibinfo{author}{\bibfnamefont{S.}~\bibnamefont{Satpathy}},
  \bibinfo{author}{\bibfnamefont{Z.~S.} \bibnamefont{Popovic}},
  \bibnamefont{and} \bibinfo{author}{\bibfnamefont{F.~R.}
  \bibnamefont{Vukajlovic}}, \bibinfo{journal}{Phys.\ Rev.\ Lett.}
  \textbf{\bibinfo{volume}{76}}, \bibinfo{pages}{960} (\bibinfo{year}{1996}).

\bibitem[{\citenamefont{Okimoto et~al.}(1997)\citenamefont{Okimoto, Katsufuji,
  Ishikawa, Arima, and Tokura}}]{oki}
\bibinfo{author}{\bibfnamefont{Y.}~\bibnamefont{Okimoto}},
  \bibinfo{author}{\bibfnamefont{T.}~\bibnamefont{Katsufuji}},
  \bibinfo{author}{\bibfnamefont{T.}~\bibnamefont{Ishikawa}},
  \bibinfo{author}{\bibfnamefont{T.}~\bibnamefont{Arima}}, \bibnamefont{and}
  \bibinfo{author}{\bibfnamefont{Y.}~\bibnamefont{Tokura}},
  \bibinfo{journal}{Phys.\ Rev.\ B} \textbf{\bibinfo{volume}{55}},
  \bibinfo{pages}{4206} (\bibinfo{year}{1997}).

\bibitem[{\citenamefont{Nadgorny et~al.}(2001)}]{nad}
\bibinfo{author}{\bibfnamefont{B.}~\bibnamefont{Nadgorny}}
  \bibnamefont{et~al.}, \bibinfo{journal}{Phys.\ Rev.\ B}
  \textbf{\bibinfo{volume}{63}}, \bibinfo{pages}{184433}
  (\bibinfo{year}{2001}).

\bibitem[{\citenamefont{Worledge and Geballe}(2000)}]{geb}
\bibinfo{author}{\bibfnamefont{D.~C.} \bibnamefont{Worledge}} \bibnamefont{and}
  \bibinfo{author}{\bibfnamefont{T.~H.} \bibnamefont{Geballe}},
  \bibinfo{journal}{Appl.\ Phys.\ Lett.} \textbf{\bibinfo{volume}{76}},
  \bibinfo{pages}{900} (\bibinfo{year}{2000}).

\bibitem[{\citenamefont{Park et~al.}(1998)\citenamefont{Park, Vescovo, Kim,
  Kwon, Ramesh, and Venkatesan}}]{par}
\bibinfo{author}{\bibfnamefont{J.-H.} \bibnamefont{Park}},
  \bibinfo{author}{\bibfnamefont{E.}~\bibnamefont{Vescovo}},
  \bibinfo{author}{\bibfnamefont{H.-J.} \bibnamefont{Kim}},
  \bibinfo{author}{\bibfnamefont{C.}~\bibnamefont{Kwon}},
  \bibinfo{author}{\bibfnamefont{R.}~\bibnamefont{Ramesh}}, \bibnamefont{and}
  \bibinfo{author}{\bibfnamefont{T.}~\bibnamefont{Venkatesan}},
  \bibinfo{journal}{Nature} \textbf{\bibinfo{volume}{392}},
  \bibinfo{pages}{794} (\bibinfo{year}{1998}).

\bibitem[{\citenamefont{Pickett and Singh}(1996)}]{pic}
\bibinfo{author}{\bibfnamefont{W.~E.} \bibnamefont{Pickett}} \bibnamefont{and}
  \bibinfo{author}{\bibfnamefont{D.~J.} \bibnamefont{Singh}},
  \bibinfo{journal}{Phys.\ Rev.\ B} \textbf{\bibinfo{volume}{53}},
  \bibinfo{pages}{1146} (\bibinfo{year}{1996}).

\bibitem[{\citenamefont{Imada et~al.}(1998)\citenamefont{Imada, Fujimori, and
  Tokura}}]{Ima}
\bibinfo{author}{\bibfnamefont{M.}~\bibnamefont{Imada}},
  \bibinfo{author}{\bibfnamefont{A.}~\bibnamefont{Fujimori}}, \bibnamefont{and}
  \bibinfo{author}{\bibfnamefont{Y.}~\bibnamefont{Tokura}},
  \bibinfo{journal}{Rev.\ Mod.\ Phys.} \textbf{\bibinfo{volume}{70}},
  \bibinfo{pages}{1039} (\bibinfo{year}{1998}).

\end{thebibliography}
\end{document}